\documentclass[%
 aps,
 pra,
 reprint,
 amsmath,
 amssymb,
 showkeys,
 groupedaddress,
 eprint,
 longbibliography,
 nofootinbib,
 nobibnotes,
]{revtex4-1}

\usepackage[utf8]{inputenc}
\usepackage{graphicx}
\usepackage{bm}
\usepackage{xcolor}
\usepackage{enumerate}
\usepackage{amsfonts, amsmath, amssymb}
\usepackage[english]{babel}
\usepackage{dsfont}
\usepackage{placeins}
\usepackage{comment}

\usepackage{hyperref}
\hypersetup{
	colorlinks = true,
	linkcolor = blue,
	citecolor = blue,
	urlcolor = blue,
	filecolor = black,
	linktoc = all,
}

\newcommand{\ent}{\qc}
\newcommand{\VN}{\textsc{vn}}
\newcommand{\dd}{\delta}

\newcommand{\HH}{\mathcal{H}}
\newcommand{\CC}{\mathcal{C}}
\newcommand{\UU}{\mathcal{U}}
\newcommand{\one}{\mathds{1}}
\newcommand{\tr}{\textrm{tr}}
\newcommand{\ket}[1]{\left| #1 \right\rangle}
\newcommand{\bra}[1]{\left\langle #1 \right|}
\newcommand{\Phat}{\hat{P}}
\newcommand{\Qhat}{\hat{Q}}
\newcommand{\ketbra}[1]{\ket{#1}\!\bra{#1}}
\renewcommand{\P}{\hat{P}}
\newcommand{\bits}{{\rm bits}}
\newcommand{\bit}{{\rm bit}}
\newcommand{\qc}{\textsc{qc}}
\newcommand{\req}{\textsc{req}}

\begin{document}


\title{Quantum correlation entropy}

\author{Joseph Schindler}
\email{jcschind@ucsc.edu}
\affiliation{SCIPP and Department of Physics, University of California, Santa Cruz, California 95064, USA}

\author{Dominik \v{S}afr\'{a}nek}
\email{dsafrane@ucsc.edu}
\affiliation{SCIPP and Department of Physics, University of California, Santa Cruz, California 95064, USA}

\author{Anthony Aguirre}
\affiliation{SCIPP and Department of Physics, University of California, Santa Cruz, California 95064, USA}


\date{April 2020}

\begin{abstract}
We study quantum coarse-grained entropy and demonstrate that the gap in entropy between local and global coarse-grainings is a natural generalization of entanglement entropy to mixed states and multipartite systems. This ``quantum correlation entropy'' $S^\qc$ is additive over independent systems, is invariant under local unitary operations, measures total nonclassical correlations (vanishing on states with strictly classical correlation), and reduces to the entanglement entropy for bipartite pure states. It quantifies how well a quantum system can be understood via local measurements, and ties directly to non-equilibrium thermodynamics, including representing a lower bound on the quantum part of thermodynamic entropy production. We discuss two other measures of nonclassical correlation to which this entropy is equivalent, and argue that together they provide a unique thermodynamically distinguished measure.
\end{abstract}

\pacs{}
\keywords{quantum information, quantum coarse-graining, entanglement entropy, thermodynamics}

\maketitle



\section{Introduction}

Entanglement entropy is an important measure of non-local correlations in quantum systems, with uses ranging from information theory~\mbox{\cite{bennett1998quantum,cerf1998quantum,plenio2007introduction,horodecki2009quantum}}, to many-body theory~\mbox{\cite{kitaev2006topological,eisert2010colloquium,deutsch2013microscopic,zhang2015thermalization,laflorencie2016quantum}}, quantum thermodynamics~\cite{vedral2000landauer,iyoda2017fluctuation,deffner2019quantum}, quantum phase transitions~\mbox{\cite{unanyan2005entanglement,osborne2002entanglement,vidal2003entanglement}}, and the description of the holographic principle and black hole entropy~\mbox{\cite{srednicki1993entropy,ryu2006holographic,emparan2006black,nishioka2018entanglement}}.
It is defined only for pure states of a bipartite quantum system, where $S^{\rm ent}(\ketbra{\psi}) = S^{\VN}(\rho_A) = S^{\VN}(\rho_B)$, with $\ket{\psi}$ a global pure state and $\rho_A,\rho_B$ the reduced density operators in each subsystem, and with $S^{\VN}(\rho) = -\tr(\rho \log \rho)$ the von Neumann entropy.

There are various generalizations of entanglement entropy to mixed and/or multipartite states, including both measures of total nonclassical correlation~\cite{modi2012classical,bera2018quantum,adesso2016measures}, and measures of entanglement~\cite{plenio2007introduction,horodecki2009quantum}. Measures of \textit{total nonclassical correlation} (synonymously here, \textit{quantum correlation}) quantify failure to have strictly classical correlations, while measures of \textit{entanglement} quantify failure to be separable (see~\cite{modi2010unified,piani2011all} for further comparison). In bipartite pure states, where ``entanglement entropy'' is defined, the two are equivalent.%
\footnote{If a state is strictly classically correlated then it is separable. Some separable states have non-classical correlations and can exhibit inherently quantum features~\cite{modi2010unified,piani2011all}. Thus total nonclassical correlation is more general than, but includes, entanglement. Sometimes \mbox{``discord''} is used synonymously with ``total nonclassical correlation,'' but we will reserve \textit{discord} to refer to measures based on mutual information difference, like the original discord~\cite{ollivier2002quantum}.}

Many particular measures have been defined, often motivated by characterizing the usefulness of states in performing information tasks, such as quantum communication~\mbox{\cite{gisin2002quantum,pirandola2019cryptography,pirandola2015teleportation,pirandola2017fundamental,xu2019secure}}, metrology~\mbox{\cite{giovannetti2006quantum,giovannetti2011advances,Szczykulska2016multi,deffner2017quantum,pirandola2018advances}}, or computation~\mbox{\cite{nielsen2010quantum,Steane1998quantum,kjaergaard2019superconducting}}, where both quantum correlation and entanglement can be key resources~\mbox{\cite{datta2008quantum,lanyon2008experimental,piani2008local,piani2011all}}. 

These usually either quantify some type of utility (\textit{e.g.}~quantum deficit~\cite{horodecki2005local}, distillable entanglement~\cite{bennett1996concentrating,rains1999rigorous}, entanglement of formation~\cite{bennett1996mixed}, entanglement cost~\cite{hayden2001asymptotic}), or distance from some distinguished set of states (\textit{e.g.}~relative entropy of quantumness~\cite{groisman2007quantumness}, relative entropy of entanglement~\cite{vedral1997quantifying}). But they do not retain a clear interpretation as \emph{entropy}, in the sense of statistical mechanics.

Meanwhile, there are also many related but distinct notions of entropy. These range from the classical Gibbs and quantum von Neumann entropies, which are informational measures of the inherent uncertainty in a state, to ``microstate-counting entropies'' such as the classical Boltzmann entropy, and ultimately to the thermodynamic entropy and its application to heat and work.

In this context it is well known that the relationship between informational entropies and thermodynamic entropy is related to the concept of coarse-graining, as is the case with classical Boltzmann entropy. Recently, a precise formulation of coarse-graining in quantum systems, which was originally discussed by von Neumann~\cite{vonNeumann1948},%
\footnote{And see~\cite{strasberg2020heat} for a detailed account of the history of this idea.}
has been revived~\cite{Safranek2019a,Safranek2019b} and shown to provide a comprehensive framework for connecting quantum entropies to thermodynamics~\cite{Safranek2019a,Safranek2019b,safranek2019classical,strasberg2019entropy,strasberg2020heat,strasberg2020dissipation}. A key aspect of this connection is that while coarse-graining a system in energy provides a relation to equilibrium thermodynamic entropy,
non-equilibrium thermodynamic entropy relates to {\em local} energy coarse-grainings. 

In this article we study local coarse-grainings more generally, and find that there is a gap in entropy between local and global coarse-grainings that is a natural generalization of entanglement entropy to mixed and multipartite systems. This \textit{quantum correlation entropy} $S^\qc$ (``quarrelation entropy'', for short)---defined as the difference between the infimum local and global coarse-grained entropies---has both informational, statistical, and thermodynamic interpretations. As we will see, it quantifies the uncertainty of local measurements, and directly contributes to non-equilibrium thermodynamic entropy.

The quantum correlation entropy, as well as being a statistical/thermodynamic entropy, is a measure of total nonclassical correlations. As discussed further in Sec.~\ref{sec:equivalence}, it is equal to two other such measures: the relative entropy of quantumness~\mbox{\cite{modi2010unified,saitoh2008nonclassical,groisman2007quantumness,brodutch2018quantumness,horodecki2005local}} (a~measure of distance from the set of classically correlated states), and the zero-way quantum deficit~\cite{oppenheim2002thermodynamical,horodecki2003local,horodecki2005local} (a measure of work extractable by certain local operations). The equivalence of these measures, each with quite different meanings, suggests that together they have a general and distinguished role. With this in mind, this paper aims to provide a self-contained treatment in terms of the statistical mechanics of coarse-graining.

Entanglement entropy, by its usual definition, is defined only for bipartite pure states, where entanglement and total nonclassical correlation are equivalent. The generalization $S^\qc$ can be nonzero on separable states, but is zero precisely on classically correlated states. This clarifies that it measures total nonclassical correlation, not entanglement. In contrast, no entanglement measure is known to arise from statistical mechanics.%
\footnote{This raises a subtle point. The term ``entanglement entropy'' suggests that ``entanglement'' (\textit{i.e.}~non-separability) and ``entropy'' are closely related. This seems to be true \textit{only} in the special case of bipartite pure states. More generally statistical mechanical entropy appears (based on this work) tied to quantum correlations, \textit{not} entanglement, when the two are inequivalent.}

Analysis in terms of coarse-graining leads to a distinction between three types of entropy:%
\footnote{With this distinction the terms ``von Neumann entropy'' and ``entanglement entropy'' should not be applied interchangeably. While it is true that von Neumann entropy may arise in a system (say, a joint system described by $\rho_{AB}$) \textit{because of} its entanglement with some external system (say, system~$C$), this is a fundamentally different concept than quantum correlation entropy (\textit{i.e.}~entanglement entropy) \textit{within} the system (that is, between $A$ and~$B$).}
\begin{itemize}
    \item \textit{von Neumann entropy} is inherent to the state $\rho$, and quantifies fundamental uncertainty in a system due to being in a mixed state;
    \item \textit{quantum correlation entropy} (equivalently, where it is defined, \textit{entanglement entropy}) depends on a partition into subsystems, and quantifies the additional uncertainty in a multipartite system if one can only make subsystem-local measurements;
    \item \textit{coarse-grained entropy} depends on a division of the state space into macrostates, and quantifies uncertainty associated with describing a system in terms of these macrostates.
\end{itemize}

The first two each contribute to the third: the entropy of any possible coarse-graining is bounded below by von Neumann entropy, while the entropy of any \emph{local} coarse-graining is bounded below by the sum of von Neumann and quantum correlation entropies.%
\footnote{Recalling its connection to local energy coarse-graining, this sum is then a bound on non-equilibrium thermodynamic entropy.}

In this way quantum correlation entropy provides a key piece to a unified treatment of quantum statistical/thermodynamic entropy, along with a direct link to important measures in quantum information theory.

\section{Quantum correlation entropy from coarse-grained entropy}

In the theory of quantum coarse-grained entropy~\cite{vonNeumann1948,Safranek2019a,Safranek2019b,safranek2019classical,strasberg2020heat}, a coarse-graining $\CC=\{\Phat_i\}$ is a collection of Hermitian ($\Phat_i^\dag = \Phat_i$) orthogonal projectors ($\Phat_i \Phat_j = \Phat_i \, \dd_{ij}$) forming a partition of unity ($\sum_i \Phat_i =  \one$). A coarse-graining is the set of outcomes of some projective measurement. Each subspace generated by $\Phat_i$ is called a ``macrostate.''

Given a coarse-graining $\CC$ the ``coarse-grained entropy'' (or ``observational entropy'') of a density operator $\rho$ is
\begin{equation}
\label{eqn:obs-ent}
    S_{\CC}(\rho) = -\sum_i p_i \log \left( \frac{p_i}{V_i} \right),
\end{equation}
where $p_i = \tr(\Phat_i \rho)$ is the probability to find $\rho$ in each macrostate, and $V_i = \tr(\Phat_i)$ is the volume of each macrostate. The coarse-grained entropy is defined both in and out of equilibrium, obeys a second law, and (with a properly chosen coarse-graining) is equal to thermodynamic entropy in appropriate cases~\cite{Safranek2019a,Safranek2019b,safranek2019classical,strasberg2020heat,strasberg2020dissipation,strasberg2019entropy,lent2019quantum,gemmer2014entropy,goldstein2019gibbs}.

One way to specify a coarse-graining is via the spectral decomposition of an observable operator $\hat{Q} = \sum_q q \, \Phat_q$, with associated coarse-graining $\CC_{\hat{Q}} = \{ \Phat_q \}$. If $\Qhat$ has a full spectrum of distinct eigenvalues, then $S_{\CC_{\Qhat}}(\rho)$ is merely the Shannon entropy of measuring $\Qhat$. On the other hand $\Qhat$ may have larger eigenspaces. If $\rho$ has a definite value $q$ then $S_{\CC_{\Qhat}}(\rho)$ is the log of the dimension of the $q$ eigenspace of $\Qhat$ (\textit{i.e.} the volume of the $q$ macrostate), a quantum analog of the Boltzmann entropy. Evidently the coarse-grained entropy provides a quantum generalization of both the Shannon and Boltzmann entropies of a measurement, and represents the uncertainty an observer making measurements assigns to the system.

Given a density operator $\rho$, the minimum value of coarse-grained entropy, minimized over all coarse-grainings $\CC$, is
\begin{equation}
\label{eqn:vn-is-min}
    \inf_\CC \Big(S_{\CC}(\rho) \Big) = S_{\CC_{\rho}}(\rho) = S^{\VN}(\rho),
\end{equation}
the von Neumann entropy~\cite{vonNeumann1948,Safranek2019a,Safranek2019b}. The second equality states that the von Neumann entropy $S^{\VN}(\rho) = - \tr(\rho \log \rho)$ is equal to the coarse-grained entropy $S_{\CC_{\rho}}(\rho)$ in the coarse-graining $\CC_\rho$ consisting of eigenspaces of $\rho$.  Thus (\ref{eqn:vn-is-min}) expresses that no measurement can be more informative than measuring the density matrix itself.

Now consider an arbitrary multipartite system $AB\ldots C$, whose Hilbert space is the tensor product $\HH = \HH_A \otimes \HH_B \otimes \ldots \otimes \HH_C$.

In this background one can consider a subclass of coarse-grainings, the ``local'' coarse-grainings, defined by 
\begin{equation}
\label{eqn:product-coarse-graining}
    \CC_A \otimes \CC_B \otimes \ldots \otimes \CC_C = \{ \Phat^A_l \otimes \Phat^B_m \otimes \ldots \otimes \Phat^C_n \},
\end{equation}
where 
$\CC_A= \{ \Phat^A_l \}$
is a coarse-graining of $A$, and so on for the other subsystems. These are precisely the coarse-grainings describing local measurements (\textit{i.e.} consisting of only local operators). Applying the definition (\ref{eqn:obs-ent}) in such a coarse-graining yields the entropy
\begin{equation}
\label{eqn:obsent-prod}
    S_{\CC_A \otimes \ldots \otimes \CC_C}(\rho) = 
    -\sum_{lm \ldots n} p_{lm \ldots n} \log \left( \frac{p_{lm \ldots n}}{V_{lm \ldots n}} \right),
\end{equation}
where $p_{lm \ldots n} = \tr(\Phat^A_l \otimes \Phat^B_m \otimes \ldots \otimes \Phat^C_n \,\rho)$ are the probabilities to find the system in each macrostate, and $V_{lm \ldots n} = \tr(\Phat^A_l \otimes \Phat^B_m \otimes \ldots \otimes \Phat^C_n)$ are the volumes of each macrostate.

One can now ask: what is the minimum entropy of any set of local measurements? That is, what is the infimum value
\begin{equation}
    \inf_{\CC = \CC_A \otimes \ldots \otimes \CC_C} \Big(S_{\CC}(\rho) \Big)
\end{equation}
of the coarse-grained entropy, if the minimum is taken \textit{only over local coarse-grainings}?

There are two possibilities. Either the minimum value $S^{\VN}(\rho)$ can be saturated by local coarse-grainings, or it cannot. Which of these is the case depends on the density matrix. If the minimum fails to be saturated, then there is an entropy gap $\Delta S$ above the minimum which is inherent to \textit{any} local measurements. 

A natural question is then: how is this entropy gap, associated with restricting to local coarse-grainings, related to entanglement entropy? Two observations provide a foundation for answering this question. The first, quite nontrivial, observation is that the entropy gap is equal to the entanglement entropy for bipartite pure states (see Property Ia). The second is that the entropy gap is zero for any product state (see Property II). These facts suggest that entanglement entropy should in general be identified with this entropy gap. The aim of this article is to make precisely that identification and show that it leads to an intuitive and useful framework.

The observations above motivate the definition
\begin{equation}
\label{eqn:ee}
    S^{\ent}_{AB \ldots C}(\rho) \equiv 
    \inf_{\CC = \CC_A \otimes \ldots \otimes \CC_C} \Big(S_{\CC}(\rho) \Big) - S^{\VN}(\rho)
\end{equation}
of the quantum correlation (quarrelation) entropy $S^{\ent}_{AB \ldots C}(\rho)$. The subscript denotes the partition into subsystems, allowing various partitions of the same system.

In other words, quarrelation entropy is the difference in coarse-grained entropy between the best possible local coarse-graining and the best possible global coarse-graining. This definition can be evaluated exactly for a variety of states using the properties introduced below, and can also be implemented numerically.

\section{Properties}

The quantum correlation (quarrelation) entropy $S^{\ent}$, defined by Eq.~(\ref{eqn:ee}), has the following properties. Proofs are given in the appendix.%
\footnote{Note that properties (I), (IIa), (VII), and equations (12), (16), have appeared before in the literature in the context of equivalent measures (see section \ref{sec:equivalence} for further discussion). Also Bravyi~\cite{bravyi2003entanglement} has evaluated an equivalent measure on the so-called determinant and hexacode pure states.}

(Ia) A bipartite system $AB$ in a pure state $\rho = \ket{\psi}\!\bra{\psi}$, with reduced densities $\rho_A$ and $\rho_B$ in the $A$ and $B$ subsystems, has quarrelation entropy
\begin{equation}
    S^{\ent}_{AB}(\rho) = S^{\VN}(\rho_A) = S^{\VN}(\rho_B).
\end{equation}
This is equal to the usual entanglement entropy.

(Ib) More generally, for any multipartite state of the special (``maximally correlated''~\cite{rains1999bound,oppenheim2002thermodynamical}) form
\begin{equation}
\label{eqn:max-cor}
    \rho = \sum_{ij} \sigma_{ij} \ket{a_i b_i \ldots c_i} \bra{a_j b_j \ldots c_j},
\end{equation}
where $\sigma_{ij}$ are complex coefficients and $\ket{a_l},\ket{b_m}, \ldots, \ket{c_n}$ are orthonormal bases for the $A,B,\ldots,C$ subsystems, the quarrelation entropy is
\begin{equation}
    S^{\ent}_{AB\ldots C}(\rho) = \Big( -\sum_i \sigma_{ii} \log \sigma_{ii} \Big) - S^{\VN}(\rho).
\end{equation}
Note that for $\rho$ to be a state requires $\sigma_{ij}=\sigma_{ji}^*$ and $\sum_i \sigma_{ii}=1$. These states include all pure states of the form $\ket{\psi} = \sum_k \alpha_k \ket{a_k b_k \ldots c_k}$, and thus all bipartite pure states by virtue of the Schmidt decomposition. The infimum defining quarrelation entropy is achieved by coarse-graining in the $\ket{a_l b_m \ldots c_n}$ basis.

(IIa) In finite dimensions, $S^{\ent}_{AB \ldots C}(\rho) = 0$ if and only if $\rho$ is a classically correlated state---that is, if there exists a locally orthonormal product basis diagonalizing $\rho$. Explicitly, \textit{classically correlated states} (sometimes called ``strictly classically correlated'') are those that can be put in the form
\begin{equation}
\label{eqn:classical-state}
\rho = \sum_{l m\ldots n} p_{lm\ldots n} \ketbra{a_l b_m \ldots c_n}
\end{equation}
where $\ket{a_l}, \ldots, \ket{c_n}$ form orthonormal bases in $A, \dots, C$, and $p_{lm \ldots n}$ form a set of real probabilities. That these are the states with strictly classical correlations has been studied extensively~\cite{ollivier2002quantum,oppenheim2002thermodynamical,horodecki2005local,groisman2007quantumness,luo2008using,li2008classical,modi2010unified,piani2011all}. Classically correlated states include all product states, and form a strict subset of the separable states. 

(IIb) In general (finite or infinite dimensions), $\rho$~is a classically correlated state if and only if both $\inf_{\CC = \CC_A \otimes \ldots \otimes \CC_C} S_{\CC}(\rho)$ is realized as a minimum and $S^{\ent}_{AB \ldots C}(\rho) = 0$. In finite dimensions the infimum is always realized.

(III) For any local coarse graining $\CC_A \otimes \ldots \otimes \CC_C$,
\begin{equation}
\label{eq:lower_bound}
    S_{\CC_A \otimes \ldots \otimes \CC_C}(\rho)\geq S^{\VN}(\rho) + S^{\ent}_{AB \ldots C}(\rho).
\end{equation}
That is, any observer who can make only local measurements observes at least as much uncertainty as the inherent uncertainty in the joint state (the von Neumann entropy) plus an additional contribution (the quarrelation entropy) due to their inability to make a nonlocal joint measurement.

(IV) In general $0 \leq S^{\ent}_{AB\ldots C}(\rho) \leq \log \dim \HH - S^{\VN}(\rho)$. Additional bounds can be computed directly from local von Neumann entropies. A family of lower bounds is given by
\begin{equation}
    S^{\ent}_{AB\ldots C}(\rho)
        \geq
         S^{\VN}(\rho_{\rm loc})  - S^{\VN}(\rho),
\end{equation}
where $\rho_{\rm loc}$ is any local reduced density matrix obtained by tracing out some of the subsystems. An upper bound is given by
\begin{equation}
\label{eqn:upper-bound}
    S^{\ent}_{AB\ldots C}(\rho)
        \leq
        \Big(\displaystyle\sum_{X} S^{\VN}(\rho_{X}) \Big)
        - S^{\VN}(\rho),  
\end{equation}
where $X\in\{A,B,\ldots,C\}$ is an index summing over all the subsystems, with $\rho_X$ the reduced density in each one.

(V) If $\HH = \HH_A \otimes \HH_B$ and $\HH_B = \HH_{B_1} \otimes \HH_{B_2}$ then for a fixed $\rho$ on $\HH$,
\begin{equation}
    S^{\ent}_{A B_1 B_2}(\rho) \geq S^{\ent}_{A B}(\rho).
\end{equation}
That is, further splitting up the system into smaller subsystems can only increase quarrelation entropy.

(VI) If $\HH = \HH_A \otimes \HH_B$ and $\HH_B = \HH_{B_1} \otimes \HH_{B_2}$ then
\begin{equation}
    \label{eqn:additivity}
    S^{\ent}_{A B_1 B_2}(\rho_A\otimes \rho_B) = S^{\ent}_{B_1 B_2}(\rho_B).
\end{equation}
If also $\HH_A = \HH_{A_1} \otimes \HH_{A_2}$ then
\begin{equation}
    S^{\ent}_{A_1 A_2 B_1 B_2}(\rho_A\otimes \rho_B) = S^{\ent}_{A_1 A_2}(\rho_A)+S^{\ent}_{B_1 B_2}(\rho_B).
\end{equation}
That is, $S^{\ent}$ is additive on independent systems.

(VII) $S^{\ent}_{AB \ldots C}(\rho)$ is invariant under local unitary operations. That is, if $\Tilde{\rho} = (U_A \otimes \ldots \otimes U_C) \, \rho \, (U_A^\dag \otimes \ldots \otimes U_C^\dag)$, then $S^{\ent}_{AB\dots C}(\Tilde{\rho}) = S^{\ent}_{AB\dots C}(\rho)$, where $U$ are local unitaries.

\section{Relationship to subsystem entropies and mutual information}

In order to understand the quantum correlation (quarrelation) entropy it is instructive to see \textit{how} the entropy of a local coarse-graining is minimized, by considering the identity
\begin{equation}
\label{eqn:product-fomula}
    S_{\CC_A \otimes \ldots \otimes \CC_C}(\rho) = 
    \Big( \sum_{X} S_{\CC_X}(\rho_X) \Big)
    - I_{\CC_A \otimes \ldots \otimes \CC_C}(\rho),
\end{equation}
where $X \in \{A,B,\ldots,C \}$ labels the subsystems, with $\rho_X$ the reduced density in each one, and
\begin{equation}
\label{eqn:mutual-info}
    I_{\CC_A \otimes \ldots \otimes \CC_C}(\rho) \equiv \sum_{lm \ldots n} p_{lm \ldots n} \log \left( \frac{p_{lm \ldots n}}{p^A_l p^B_m \ldots p^C_n} \right)
\end{equation}
is the mutual information of the joint measurement. The~$p^A_l \equiv \sum_{m\dots n}p_{lm\dots n}=\tr(\Phat^A_l\rho_A)$ and so on are marginal probabilities, and subadditivity of Shannon entropy implies $I \geq 0$.

In computing $S^\qc$ one might hope to minimize the subsystem entropies $S_{C_X}$  while maximizing $I_{\CC_A \otimes \ldots \otimes \CC_C}$ in~(\ref{eqn:product-fomula}). These extrema cannot, in general, be achieved simultaneously, so an optimal coarse-graining must obtain some balance of these contributions.

Pure states of the form $\ket{\psi} = \sum_k \alpha_k \ket{a_k b_k \ldots c_k}$ (\textit{cf.}~property (Ib)) provide a special case where the subsystem entropies and mutual information can be simultaneously extremized. In the optimal coarse-graining, assuming $N$ subsystems, one then finds 
$\sum_{X} S_{C_X}(\rho_X) = N S_0$ 
and 
$I = (N-1) S_0$, where 
\mbox{$S_0 = -\sum_k |\alpha_k|^2 \log \left( |\alpha_k|^2 \right)$}.
Subtracting these two contributions leads in this special case to
\begin{equation}
\label{eqn:special-cases}
    S^{\ent}_{AB\ldots C}(\rho) = S^{\VN}(\rho_A) = \ldots = S^{\VN}(\rho_C) = S_0,
\end{equation}
an equality which could be somewhat misleading, since in general the quarrelation entropy and subsystem von Neumann entropies will not be equal.

\section{Examples}
\label{sec:examples}

To demonstrate calculability we exhibit two simple examples of some relevance to the literature. 
Example (A) compares ``two Bell pair'' versus GHZ entanglement in different partitions, relevant to genuine multipartite nonlocality~\cite{schmid2020why,navascues2020genuine,walter2016multipartite}.
Example (B) considers a prototypical ``separable but not classically correlated'' state, relevant to local indistinguishablity~\cite{bennett1999quantum}.

(A) In a 4-partite system labelled $A_1 \otimes A_2 \otimes B_1 \otimes B_2$, define $\ket{\phi_{GHZ}}=(\ket{0000}+\ket{1111})/\sqrt{2}$ and  $\ket{\phi_{2Bell}}=\ket{\phi^{+}}_{A_1 B_1} \otimes \ket{\phi^{+}}_{A_2 B_2}$, where $\ket{\phi^{+}}=(\ket{00}+\ket{11})/\sqrt{2}$, each with density $\rho=\ketbra{\phi}$. By properties (I,VI) above, we find for the 4-partite case $S^{\ent}_{A_1 A_2 B_1 B_2}(\rho_{2Bell}) = 2~\bits$, while in two bipartite cases $S^{\ent}_{(A_1 \cup B_1)(A_2 \cup B_2)}(\rho_{2Bell}) = 0$ and $S^{\ent}_{AB}(\rho_{2Bell}) = 2~\bits$. Meanwhile $S^{\ent}(\rho_{GHZ})=1~\bit$ in all these partitions.

(B) In a bipartite system $A \otimes B$ define $\rho = \frac{1}{2} (\ketbra{00}+\ketbra{1+})$, where $\ket{+}=(\ket{0}+\ket{1})/\sqrt{2}$. Properties (IV,II) provide an analytical bound $\alpha \geq S^{\ent}_{AB}(\rho) > 0$ (where $\alpha \approx 0.6~\bits$ is a number derived from (\ref{eqn:upper-bound})). Numerical minimization estimates $S^{\ent}_{AB}(\rho) = 0.50~\bits$. 

\section{Equivalence to other measures}
\label{sec:equivalence}

Equivalent measures to the entropy considered here have arisen with various motivations and in various guises throughout the literature. The first seems to have been considered (in the special case of pure states) by Bravyi~\cite{bravyi2003entanglement} as a minimal Shannon entropy of measurement outcomes. The motivation was essentially similar to here, only lacking the connection to coarse-graining and statistical mechanics. This was generalized to mixed states by SaiToh~\textit{et.~al.}~\cite{saitoh2008nonclassical}, though without reference to Bravyi. In between those studies the concept of quantum deficit was introduced by Oppenheim~\textit{et.~al.}~\cite{oppenheim2002thermodynamical} and in subsequent studies~\cite{horodecki2003local,horodecki2005local} the zero-way deficit was shown (implicitly) to be equal to the measure of Bravyi and SaiToh~\textit{et.~al.}~and also (explicitly) to the relative entropy distance to classically correlated states. That distance was then proposed as an important measure of nonclassicality in its own right by Groisman~\textit{et.~al.}~\cite{groisman2007quantumness}, who called it relative entropy of quantumness, and systematically related to other relative entropy based measures by \mbox{Modi~\textit{et.~al.}~\cite{modi2010unified}}, who called it the relative entropy of discord. More recently, similar quantities have appeared related to quantum coherence~\cite{baumgratz2014quantifying,levi2014quantitative,streltsov2017quantum,radhakrishnan2016distribution,tan2016unified,yao2017comment,bu2017distribution,wang2017relating,xi2018coherence,kraft2018genuine}, where relative entropy of quantumness is the minimum over local bases of relative entropy of coherence.

Here we are interested in the equivalence of three quantities. The zero-way quantum deficit, which measures a difference in work extractable by local versus global operations, is defined by~\cite{horodecki2005local}
\begin{equation}
    \Delta^{\emptyset}(\rho) = \inf_{\Lambda \in {\rm CLOCC}^{\emptyset}} [S(\rho'_A) + \ldots + S(\rho'_C)] - S(\rho),
\end{equation}
where $\Lambda$ is a zero-way CLOCC operation (see~\cite{horodecki2005local}), $\rho' = \Lambda(\rho)$, and $\rho_A'=\tr_{B\dots C}(\rho')$ and so on are the reduced densities. The relative entropy of quantumness (also known as the relative entropy of discord~\cite{modi2010unified}) measures distance to the nearest classically correlated state, and is defined by~\cite{groisman2007quantumness,piani2011all}
\begin{equation}
\label{eqn:sreq}
    S^{\req}(\rho) = \inf_{\chi \in \chi_c} S(\rho \,|| \, \chi),
\end{equation}
where $\chi_c$ is the set of all classically correlated states as defined by (\ref{eqn:classical-state}) and $S(\rho \,|| \, \chi) = \tr(\rho \log \rho - \rho \log \chi)$ is the quantum relative entropy. And $S^\ent(\rho)$ is defined by (\ref{eqn:ee}).

It is well known that $\Delta^\emptyset = S^{\req}$~\cite{horodecki2005local,modi2012classical}. It is easy to also show $S^\ent = S^{\req}$. By Thm.~3 of~\cite{Safranek2019b}, every coarse-graining can be refined to rank-1 without increasing coarse-grained entropy, so that (\ref{eqn:ee}) can be rewritten as an infimum over \mbox{rank-1} local projectors. Then an application of Lemma 1 from~\cite{horodecki2005local} leaves the composition of two infima which combine to the one in (\ref{eqn:sreq}) above. Thus $\Delta^\emptyset = S^{\req}= S^{\ent}$.

This measure therefore has three significant and complementary interpretations: (1) in terms of work extractable by local operations, (2) as the distance from the set of classically correlated states, and (3) as a statistical mechanical entropy related to non-equilibrium thermodynamics. Given this breadth of meaning, these three quantities seem to provide a thermodynamically distinguished measure of nonclassical correlation.

\section{Inequivalent but related measures}

In the special case where both terms of (\ref{eqn:product-fomula}) can be extremized by the same coarse-graining, $S^\qc$ becomes equal to other related measures. To exemplify this in a simple setting, consider a bipartite state $\rho_{AB}$ such that $\rho_A$, $\rho_B$, and $\rho_A \otimes \rho_B$ all have nondegenerate eigenspaces.

First note that $S^\qc$ can be related, in general, to a difference in mutual information. Defining the quantum mutual information as $I_{\rm qm} = S^{\VN}(\rho_A) + S^{\VN}(\rho_B) - S^{\VN}(\rho_{AB})$ and the classical mutual information $I_{\rm cl}$ as the supremum of (\ref{eqn:mutual-info}) over local coarse-grainings, one finds
\begin{equation}
\label{eqn:delta-I}
    S^\qc_{AB} \geq I_{\rm qm} - I_{\rm cl}.
\end{equation}
This is proved by plugging (\ref{eqn:product-fomula}) into (\ref{eqn:ee}) and applying $\inf(x-y) \geq \inf(x) - \sup(y)$.

If there exists a local coarse-graining $\CC_0$ that simultaneously infimizes the marginal term and supremizes the mutual term in (\ref{eqn:product-fomula}), then (\ref{eqn:delta-I}) becomes an equality, and $\CC_0=\CC_{\rho_A} \otimes \CC_{\rho_B}=\CC_{\rho_A \otimes \rho_B}$ is the coarse-graining in reduced density matrix eigenbases (we have used the simplifying assumption about nondegeneracy here). Then, in this special case,
\begin{equation}
\label{eqn:related-measures}
\begin{array}{rcl}
    S^\qc_{AB}(\rho_{AB}) &=& S_{\CC_{\rho_A} \otimes \, \CC_{\rho_B}}(\rho_{AB}) - S^\VN(\rho_{AB}) \\[12pt]
    &=& I_{\rm qm} - I_{\rm cl}.
\end{array}
\end{equation}
This difference of mutual information is a symmetric discord measure~\cite{rulli2011global,maziero2010symmetry,okrasa2011quantum}, and is closely related to measures of correlated coherence~\cite{tan2016unified,yao2017comment,wang2017relating,xi2018coherence,kraft2018genuine}.

But the equality (\ref{eqn:related-measures}) does not hold in general---generically there may be three distinct coarse-grainings: infimizing the marginal term, supremizing the mutual term, and infimizing their difference. Then all three quantities in~(\ref{eqn:related-measures}) are inequivalent, and only $S^{\qc}$ has a simple interpretation as minimal coarse-grained entropy. The equality in (\ref{eqn:delta-I}) does hold in most simple examples, but a counterexample to equality is given by the state in Example~(B) of Sec.~\ref{sec:examples} above, where strict inequality can be observed numerically.

\section{Discussion and Conclusions}

Consider a state $\rho$ in a multipartite system. The coarse-grained entropy of $\rho$, when minimized over all possible coarse-grainings, has a minimum given by the von Neumann entropy. But if one minimizes over only \textit{local} coarse-grainings, the minimum may be higher. This entropy gap is what we call the quantum correlation (quarrelation) entropy.

This definition treats pure and mixed states, and multipartite systems with any number of subsystems, all on equal footing. It is also a measure of total nonclassical correlation: it is equal to the zero-way quantum deficit and to the relative entropy of quantumness. Together these provide a clear interpretation: this entropy arises because no set of local measurements can reveal all information about a state with nonclassical correlations.

The given definition can be extended immediately to classical systems (described by phase space density distributions) in the context of classical coarse-grained entropy~\cite{safranek2019classical}, but in the classical case $S^{\ent}$ is always zero. This reflects that, like classically correlated quantum states (\textit{cf.} (\ref{eqn:classical-state})), the state of a classical system is exactly determined by local measurements.

In addition to measuring nonclassical correlation, this entropy has a role in thermodynamics. So far quantum coarse-grained entropy has been formally applied to non-equilibrium thermodynamics in two main scenarios.

In one scenario, Strasberg and Winter~\cite{strasberg2020dissipation,strasberg2020heat} considered a system-bath interaction where total thermodynamic entropy was identified as $S_{\CC_S \otimes \CC_{E}}$ (in the present notation), where $\CC_{E}$ is an energy coarse-graining of the bath, and $\CC_S$ is any coarse-graining of the system. This entropy was shown to be produced by non-equilibrium processes in accordance with standard thermodynamic laws. The present work indicates that one factor behind entropy production is the development of nonclassical correlations between the system and bath, and in particular that $S_{\CC_S \otimes \CC_{E}} \geq S^{\ent}_{SB}(\rho) + S^{\VN}(\rho)$. Strasberg and Winter also showed that the entropy production splits into classical and quantum parts---and comparing to (21) and (42) of~\cite{strasberg2020heat}, $S^{\qc}$ here is a lower bound on the quantum part alone. This quantum entropy production coincides with the relative entropy of coherence in the coarse-graining basis, and recent studies of nonequilibrium thermodynamics based on coherence~\cite{santos2019role,francica2019role,dolatkhah2020entropy}, and other methods~\cite{bera2017generalized}, have also led to related observations.

The other scenario considered~\cite{Safranek2019a,Safranek2019b} was thermalization in a closed isolated system with local interactions, initialized away from equilibrium. Nonequilibrium thermodynamic entropy in this case can be identified with observational entropy $S_{\CC}(\rho)$ in a local energy coarse-graining $\CC = \otimes_i \, \CC_{H_i}$, where the system is split into small but macroscopic local subsystems each with local Hamiltonian $H_i$. Starting out of equilibrium, over time this entropy dynamically approaches the expected equilibrium value (up to some corrections dependent upon finite-size effects and on the initial state)---even though the system is closed, and perhaps pure. 
Not only does this entropy dynamically equilibrate, it also has a clear interpretation when the system has only partially equilibrated~\cite{safranek2019classical,safranek2020short}.
Comparison with an equivalent classical scenario shows that this entropy increases in both situations~\cite{safranek2019classical}. Through~\eqref{eq:lower_bound}, the present work shows that in the quantum case, creation of nonclassical correlations is an extra factor that drives the entropy upwards.

In both cases, nonequilibrium thermodynamic entropy can be seen as arising from some appropriate local coarse-graining, and thus has three additive (non-negative) contributions: (1)~$S^{\VN}(\rho)$, the mixedness of the global state; (2)~$S^{\ent}(\rho)$, the entropy of nonclassical correlation between the relevant subsystems; and (3) an additional contribution depending on the specific coarse-graining relevant to the problem. 

Quantum correlation entropy thus provides useful insight into the relations between thermalization, entropy production, and nonclassical correlation, and clarifies how entanglement entropy---as a statistical mechanical entropy---generalizes to generic systems.


\begin{acknowledgments}
We thank J.~Deutsch for useful discussions, and P.~Strasberg and A.~Winter for useful comments. This research was supported by the Foundational Questions Institute (FQXi.org), of which AA is Associate Director, and by the Faggin Presidential Chair Fund.
\end{acknowledgments}


\appendix*

\section{}

This Appendix provides proofs of the properties \mbox{(I--VII)} of $S^{\ent}$ listed in the main text.

(Ia) This follows from (Ib), but we show the special case here for clarity. Every bipartite pure state can be put in the form $\ket{\psi} = \sum_k c_k \ket{a_k b_k}$, $\rho=\ketbra{\psi}$, by Schmidt decomposition. The reduced densities obey $S_0 = -\sum_k |c_k|^2 \log |c_k|^2 = S(\rho_A)=S(\rho_B)$. Evaluating in the local coarse-graining $\CC = \{ \ketbra{a_l b_m}\}$ yields $S_{\CC}(\rho) = S_0$. So by (\ref{eqn:ee}), $S^\qc_{AB}(\rho) \leq S_0$. But we can also show $S^\qc_{AB}(\rho) \geq S_0$, by considering marginal entropies, as follows. The lemma in the proof of (IV) shows that, for any local coarse-graining, $S_{\CC_A \otimes \CC_B}(\rho) \geq S_{\CC_A}(\rho_A)$. But $S_{\CC_A}(\rho_A) \geq S^\VN(\rho_A)$. Since every local coarse-graining obeys $S_{\CC_A \otimes \CC_B}(\rho) \geq S(\rho_A)$, so does the infimum, so $S^\qc_{AB}(\rho) \geq S_0$. Thus, $S^\qc_{AB}(\rho) = S_0$.

(Ib) Let $S_0 = -\sum_i \sigma_{ii} \log \sigma_{ii} - S^{\VN}(\rho)$. $\CC = \{ \ketbra{a_l} \} \otimes \ldots  \otimes \{ \ketbra{c_n} \}$ is a local coarse-graining such that $S_{\CC}(\rho) = S_0 + S^{\VN}(\rho)$. So $S^{\ent}_{AB \ldots C}(\rho) \leq  S_0$. But~(IV) with $\rho_{\rm loc}=\rho_A$ gives also $S^{\ent}_{AB \ldots C} (\rho)\geq  S_0$. Thus \mbox{$S^{\ent}_{AB \ldots C} (\rho)= S_0$}.

(IIa) Follows from (IIb).

(IIb) 
($\Rightarrow$) 
Suppose $\ket{a_l \ldots c_n}$ is a product basis diagonalizing $\rho$. Then $\CC = \{\ketbra{a_l}\} \otimes \ldots \otimes \{\ketbra{c_n}\} = \{ \ketbra{a_l \ldots c_n}\}$ is a local coarse-graining finer than $\CC_{\rho}$, which implies $S_{\CC}(\rho) = S^{\VN}(\rho)$ (Thm.~3 of \cite{Safranek2019b}), which is the infimum by (\ref{eqn:vn-is-min}).
($\Leftarrow$) 
Assume $\min_{\CC = \CC_A \otimes \ldots \otimes \CC_C} S_{\CC}(\rho)$ exists and is equal to $S^{\VN}(\rho)$. The coarse-graining $\CC_0=\{\P_l\otimes\cdots\otimes\P_n\}$ attaining the minimum is finer than $\CC_{\rho}$ (Thm.~3 of~\cite{Safranek2019b}), thus it diagonalizes $\rho$. Thus $\rho=\sum_{l\dots n}p_{l\dots n}\P_l\otimes\cdots\otimes\P_n$ where $p_{l\dots n}$ are real numbers. Writing each projector into rank-1 orthogonal projectors $\P_l=\sum_{k_l}\ketbra{a_{k_l}}$ yields the classically correlated form (\ref{eqn:classical-state}). Then $\{\ket{a_{k_l} \ldots c_{k_n}}\}$ is a product basis diagonalizing $\rho$. 
(Finite Dimensions) 
Only coarse-grainings involving rank-1 projectors need be considered in the infimum since others can be refined (Thm.~2 of~\cite{Safranek2019b}). These can be written in terms of unitary operators $U_A, \ldots, U_C$ such that
$\inf_{\CC = \CC_A \otimes \ldots \otimes \CC_C}
S_{\CC}(\rho) 
= 
\inf_{(U_A, \ldots, U_C)}
\widetilde{S}(U_A, \ldots, U_C)$
where
$\widetilde{S} = -\sum_{l \ldots n} p_{l \ldots n} \log p_{l \ldots n}$,
with 
$p_{l \ldots n} 
\equiv 
\tr( \rho \, (U_A \otimes \ldots \otimes U_C)^\dag \Phat_{l\ldots n} (U_A \otimes \ldots \otimes U_C))$
and $\Phat_{l \ldots n}$ are projectors of any rank-one local coarse-graining.
Then $\widetilde{S}: \UU_A \times \ldots \times \UU_C \to \mathbb{R}$ with $\UU_A$ the set of unitary operators on $\HH_A$, etc. If each subsystem has finite dimension then, in an appropriate topology, $\widetilde{S}(\UU_A, \ldots , \UU_C)$ is the real continuous image of a compact set, so it attains its infimum.


(III) True by definition (\ref{eqn:ee}).

(IV) The loose bounds follow immediately from (\ref{eqn:ee}) with (\ref{eqn:vn-is-min}). 
(Upper Bound) 
By (\ref{eqn:product-fomula}), since $I \geq 0$, $S_{\CC_{\rho_A} \otimes \ldots \otimes \CC_{\rho_C}}(\rho) \leq \sum_X S_{\CC_{\rho_X}}(\rho_X) = \sum_X S^{\VN}(\rho_X)$. 
But $S^{\ent}_{AB\ldots C}(\rho) \leq S_{\CC_{\rho_A} \otimes \ldots \otimes \CC_{\rho_C}}(\rho) - S^{\VN}(\rho)$ since it is the infimum. 
(Lemma)
Let $p_{lm \ldots n}$ and $V_{lm \ldots n}$ be the probabilities and volumes defining $S_{\CC_A \otimes \CC_B \otimes \ldots \otimes \CC_C}(\rho)$. 
Likewise let $q_{m \ldots n}$ and $W_{m \ldots n}$ be those defining $S_{\CC_B \otimes \ldots \otimes \CC_C}(\rho_{B \ldots C})$, where $\rho_{B \ldots C} = \tr_A(\rho)$. 
It follows that $q_{m \ldots n} = \sum_l p_{lm \ldots n}$ and $V_{m \ldots n} = \tr(\Phat^A_l) W_{m \ldots n}$. 
Thus $\frac{q_{m \ldots n}}{W_{m \ldots n}} \geq \frac{p_{lm \ldots n}}{V_{lm \ldots n}}$ for all $l,m,\ldots, n$, and since $-\log(x)$ is monotonic decreasing,
$-\sum_{lm \ldots n} p_{lm \ldots n} \log \frac{p_{lm \ldots n}}{V_{lm \ldots n}} 
\geq 
-\sum_{lm \ldots n} p_{lm \ldots n} \log \frac{q_{m \ldots n}}{W_{m \ldots n}}
=
-\sum_{m \ldots n} q_{m \ldots n} \log \frac{q_{m \ldots n}}{W_{m \ldots n}}
$.
Thus 
$S_{\CC_A \otimes \CC_B \otimes \ldots \otimes \CC_C}(\rho) \geq 
S_{\CC_B \otimes \ldots \otimes \CC_C}(\rho_{B \ldots C})$.
\mbox{(Lower Bound)} By repeated application of the lemma above,
$S_{\CC_A \otimes \ldots \otimes \CC_C \otimes \CC_D \otimes \ldots \otimes \CC_F}(\rho) 
\geq 
S_{\CC_D \otimes \ldots \otimes \CC_F}(\rho_{D \ldots F}) 
$.
But (\ref{eqn:vn-is-min}) implies
$S_{\CC_D \otimes \ldots \otimes \CC_F}(\rho_{D \ldots F})
\geq
S^{\VN}(\rho_{D \ldots F})
$.
Ordering of subsystems is irrelevant, so this is general.

(V) Any coarse-graining of the form $\CC_A \otimes \CC_{B_1} \otimes \CC_{B_2}$ is also a coarse-graining of the form $\CC_A \otimes \CC_{B}$. So one minimization strictly includes the other.

(VI) $S_{\CC_A\otimes \CC_B}(\rho_A \otimes \rho_B)=S_{\CC_A}(\rho_A)+S_{\CC_B}(\rho_B)$
since $V_{lm} =V_l V_m$ and for $\rho=\rho_A\otimes \rho_B$ also $p_{lm}=p_lp_m$.
Also $S^{\VN}(\rho_A \otimes \rho_B)=S^{\VN}(\rho_A)+S^{\VN}(\rho_B)$. Thus $S^{\ent}$ is additive since after splitting each term is infimized independently. Then (\ref{eqn:additivity}) follows from $S^{\ent}_A(\rho_A)=0$.

(VII) Write the infimum of (\ref{eqn:ee}) in terms of local unitaries as in the proof of (IIb). The local unitaries defining $\Tilde{\rho}$ are absorbed into the infimum, so the infimum is invariant. Since also $S^{\VN}(U \rho U^\dag) = S^{\VN}(\rho)$, $S^{\ent}_{AB\dots C}$ is invariant.



\bibliography{ent}

\clearpage

\end{document}